\documentstyle[11pt,newpasp,twoside,epsf]{article}
\reversemarginpar                              
\begin{document}
\title{Differential Simultaneous Imaging and Faint Companions: TRIDENT First Results from CFHT} 
\author{\sc Christian Marois, Daniel Nadeau,  Ren\'{e} Doyon, Ren\'{e} Racine}
\affil{D\'{e}partement de physique, Universit\'{e} de Montr\'{e}al, C.P. 6128, Succ. A,\\ Montr\'{e}al, QC, Canada H3C 3J7}
\author{Gordon A.H. Walker}
\affil{1234 Hewlett Place, Victoria, BC, Canada~V8S~4P7}

\begin{abstract}
We present the first results obtained at CFHT with the TRIDENT infrared camera, dedicated to the detection of faint companions close to bright nearby stars.  Its main feature is the acquisition of three simultaneous images in three wavelengths (simultaneous differential imaging) across the methane absorption bandhead at $1.6\mu$m, that enables a precise subtraction of the primary star PSF while keeping the companion signal. Thirty-five stars have been observed in two observing missions, with no detection so far. It is shown that a faint companion with a $\Delta H$ of 10 magnitudes would be detected at $0.5\arcsec$ from the primary.
\end{abstract}

\section{Introduction}
In the past few years, our group has developed the specialized infrared camera TRIDENT (Marois et al. 2000a, 2002) to search for faint companions (brown dwarfs and Jovian planets) at a projected distance of 5AU to 50AU from stars in the solar neighborhood.  This corresponds to the range of orbital distance for the massive planets of the Solar system. It is therefore of special interest to test the uniqueness of our planetary system. Ground detection of faint companions is difficult because of the atmospheric turbulence that distort the star PSF and optical aberrations that produce quasi-static PSF structures. The main feature of TRIDENT is to acquire three simultaneous images in three distinct narrow spectral bands, making possible very good atmospheric speckle and optical aberration calibration to subtract the stellar PSF. If the three simultaneous wavelengths are carefully chosen, it is possible to enhance the star/companion contrast after image combinations by selecting special spectral features that are typical of the companion and not of the star (Smith 1987; Rosenthal, Gurwell, \& Ho 1996; Racine et al. 1999; Marois et al. 2000b). In TRIDENT, the three wavelengths (1.567~$\mu$m, 1.625~$\mu$m and  1.680~$\mu$m, 1\% bandwidth) have been selected across the 1.6 $\mu$m methane absorption bandhead that is only present in the spectrum of cold ($T_{\rm{eff}}$ $<$ 1470~K, Fegley and Lodders (1996)) substellar objects. Image combinations are calculated following Marois et al. (2000b). The observations, data reduction and results are presented in section 2, 3 and 4 respectively.

\section{Observations}
Data were obtained on 2001 July 8-12 and on 2001 November 21-24, at the f/20 focus of the 3.6m CFHT adaptive optic bonnette PUEO (Rigaut et al. 1998) with TRIDENT. Flat fields and darks were obtained during the day. Three types of observing techniques were used. In July, data well inside the linear regime of the detector were acquired to test the PSF stability. Total integration time was typically 1h per target. Seeing conditions for this run were medium to good (AO strehl of 0.5 in $H$ band). In November, reference stars were acquired and instrument rotations ($+/-$ 90 degrees by step of 2 degrees) were done to calibrate or smooth non-common path aberrations discovered to be the limiting factor in the July mission data. Saturated and non-saturated images were acquired to minimize readout noise. Total integration time was 1h30 with medium seeing conditions (AO strehl from 0.15 to 0.4 in the $H$ band). In total, 35 stars have been observed during these two missions with spectral type ranging from B to M. Some stars were observed on different nights to acquire more photons and study PSF stability.

\section{Data reduction}
Dark frames having the same exposure time and number of multiple sampling readouts were subtracted from the target and flat field images. Each target image was then divided by a flat field generated by combining all flat fields obtained every night, yielding typically $10^6$ electrons per pixel. Bad and hot pixels were corrected by interpolating from nearby pixels. The 512$\times$512 pixel images at each of the three wavelengths were then extracted from the original 1024$\times$1024 pixel images. An iterative cross-correlation technique using the IDL ``rot'' procedure with cubic convolution interpolation was used for precise registering of the PSF center with respect to a common image center (precision of 0.02 pixel or 0.36~mas). A combined image (weighted by the square of the strehl ratio) for each wavelength of an object was generated by coadding all images for that object for a given night. Strehl ratios were calculated by normalizing the PSF integrated flux to unity, and dividing its central pixel by that of a theoretical PSF with the same telescope central obscuration and normalization. Strehl ratios for saturated images were taken to be the same as that of nearest previous non-saturated image. In order to combine saturated and non-saturated images, a normalization factor was calculated by measuring the flux intensities at different sections of the PSF halo. Saturated pixels (within $0.45\arcsec$ diameter) were replaced by non-saturated data. Image scaling (IDL ``rot'' procedure with cubic convolution interpolation) was then applied to correct the PSF chromatic dependence. Scaling factors obtained empirically were found to be very close to the ratio of observing wavelengths, as predicted by diffraction theory. The three simultaneous images were then normalized by measuring flux intensities in different sections of the PSF. Azymuthally averaged radial profiles were then subtracted, leaving only PSF small scale structure. Finally, image combinations were calculated from the three combined images and smoothed by a Gaussian filter with 5 pixels Full Width Half Max (same as PSF core) to average high frequency noise.

\section{Results}
The data from the July 2001 mission are limited neither by photon noise nor by speckle noise but by PSF structures that converge to a specific value for a given telescope pointing direction. Increasing the integration time beyond a few tens of seconds increases the signal to noise ratio of the PSF structure, but does not increase the detection limit. Structures seen in long integration AO corrected PSFs are not due to turbulence but to telescope and instrument optic aberrations. In theory, TRIDENT should be able to calibrate the PSF structure since each wavelength sees the same telescope and instrument aberrations. In practice, this is not exactly the case, because the three optical paths are not identical and the three final PSFs made by the TRIDENT camera are slightly decorrelated by non-common path aberrations, making impossible a perfect double difference. A typical gain of one order of magnitude in detection sensitivity is obtained with the combination of reference star subtraction and simultaneous imaging as compared with simultaneous imaging alone (see figure 1), thus confirming the quasi-static nature of the PSF structure since the reference and object were acquired at a few minute interval. Another way to minimize PSF structure effects generated after the instrument rotator (PUEO AOB and TRIDENT camera) is to smooth them by rotating the instrument. PSF attenuation obtained by subtraction after smoothing, with a strehl ratio of 0.15, is similar to that obtained with reference star PSF subtraction for data having strehl ratio of 0.5, showing that aberrations are coming mostly from PUEO and TRIDENT.

\section{Conclusion - TRIDENT new developments}
Typical attenuation in good seeing conditions are 10 magnitudes in $H$ band at $0.5\arcsec$, as good as or better than what has been done on any other telescope. Reference star offered good but not perfect PSF calibration, mainly because of the PSF structure (inside PUEO and TRIDENT) slow evolution with telescope pointing and atmospheric variations. Instrument rotation showed very good PSF smoothing, but was finally limited by primary and secondary mirror aberrations and instrument smoothed aberration residuals. Photon noise limited subtraction requires the minimization of non-common path aberrations. To increase PSF correlation, and thus subtraction performances, a new optical design is presently under evaluation that mainly consists of a better beam splitter and filters placed near the detector to minimize wavefront degradation. The goal is photon noise limited performance for long integrations (1h) without the need of a reference star or instrument rotation.\\

This work is supported in part through grants from the Natural Sciences and Engineering Research Council, Canada and from Fonds FQRNT, Qu\'{e}bec.
\clearpage

\begin{figure}
\plotone{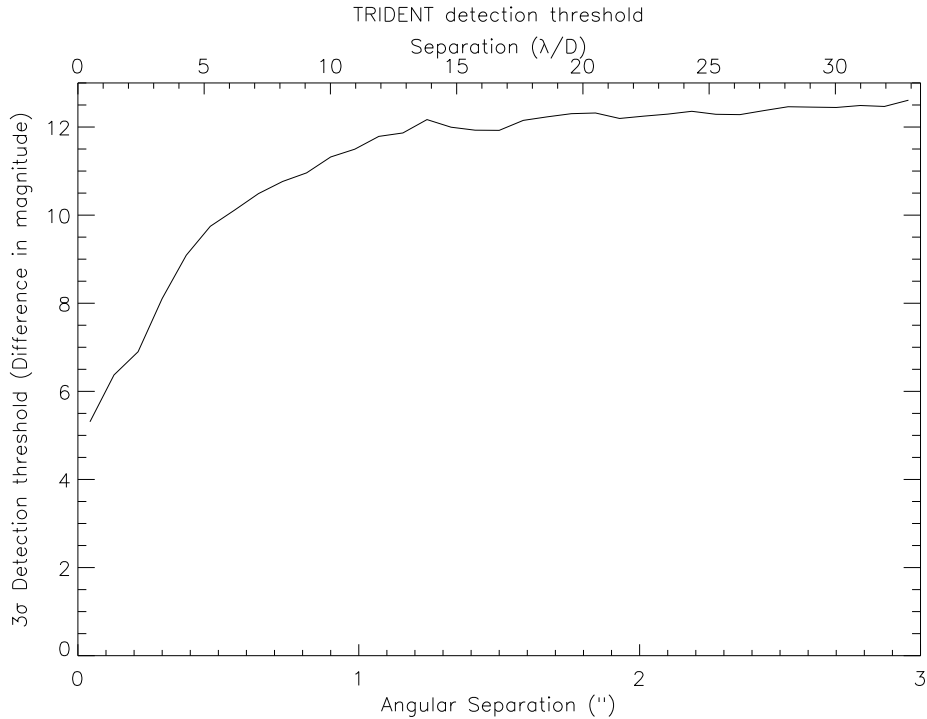}
\caption{TRIDENT detection threshold (solid line) for differential simultaneous imaging with reference star subtraction from the November 2001 CFHT $\upsilon$ And (object) and $\chi$ And (reference) data set (1h integration per target). A 10 magnitude fainter companion would be detected $0.5\arcsec$ away from the primary.}
\end{figure}


\begin{thebibliography}
\bibitem[Fegley & Lodders (1996)]{Feg96} Fegley, B.~J.,  Lodders, K. 1996, \apjl, 472, L37
\bibitem[Marois et al. (2000a)]{Marois2000a} Marois, C., Doyon, R., Racine, R. \& Nadeau, D. 2000a SPIE, 4008, 788
\bibitem[Marois et al. (2000b)]{Marois2000b} Marois, C., Doyon, R., Racine, R. \& Nadeau, D. 2000b \pasp, 112, 91
\bibitem[Marois et al. (2002)]{Marois2002} Marois, C., Nadeau, D., Doyon, R., Racine, R.  2002, in preparation
\bibitem[Smith (1987)]{Smith87} Smith W.H. 1987, \pasp, 99, 1344
\bibitem[Racine et al.(1999)]{Racine99} Racine, R., Walker, G.A.H., Nadeau, D., Doyon, R. \& Marois, C. 1999, \pasp, 111, 587
\bibitem[Rigaut et al. (1998)]{Rigaut98} Rigaut, F., Salmon, D., Arsenault, R., Thomas, J., Lai, O., Rouan, D., V{\' e}ran, J.P., Gigan, P., Crampton, D., Fletcher, J.M., Stilburn, J., Boyer, C., Jagourel, P. 1998, \pasp, 110,152
\bibitem[Rosenthal et al. (1987)]{Rosen87} Rosenthal, E.D., Gurwell, M.A., Ho, P.T.P. 1996, NATURE, 384, 243
\end{thebibliography}
\end{document}